\documentclass[sigconf,authorversion,nonacm]{acmart}

\setcopyright{cc}
\setcctype{by-sa}

\makeatletter
\renewcommand{\@copyrightpermission}{%
  {\footnotesize  This is a preprint of the paper published in the proceedings of the 2023 ACM Internet Measurement Conference (IMC ’23). The final version is available at \href{https://doi.org/10.1145/3618257.3624827}{https://doi.org/10.1145/3618257.3624827}
  } \\}
\makeatother

\usepackage[nohyperlinks, printonlyused, withpage, nolist]{acronym}

\usepackage{siunitx}  
\usepackage{enumitem}

\usepackage{tcolorbox}
\tcbset{colback=black!5,colframe=black!50}
 
\usepackage{tablefootnote}
\makeatletter
\newcommand\footnoteref[1]{\protected@xdef\@thefnmark{\ref{#1}}\@footnotemark}
\makeatother

\newcolumntype{H}{>{\setbox0=\hbox\bgroup}c<{\egroup}@{}}
\usepackage{multirow}
\usepackage{multicol}
\usepackage{threeparttable}

\PassOptionsToPackage{hyphens}{url}
\expandafter\def\expandafter\UrlBreaks\expandafter{\UrlBreaks%
	\do\a\do\b\do\c\do\d\do\e\do\f\do\g\do\h\do\i\do\j%
	\do\k\do\l\do\m\do\n\do\o\do\p\do\q\do\r\do\s\do\t%
	\do\u\do\v\do\w\do\x\do\y\do\z\do\A\do\B\do\C\do\D%
	\do\E\do\F\do\G\do\H\do\I\do\J\do\K\do\L\do\M\do\N%
	\do\O\do\P\do\Q\do\R\do\S\do\T\do\U\do\V\do\W\do\X%
	\do\Y\do\Z}

\usepackage{microtype}

\usepackage[nomessages]{fp}

\usepackage{numprint}

\usepackage{xspace}

\definecolor{cbone}  {HTML}{005B94} %
\definecolor{cbtwo}  {HTML}{FF800E} %
\definecolor{cbthree}{HTML}{ABABAB} %
\definecolor{cbfour} {HTML}{595959} %
\definecolor{cbfive} {HTML}{5F9ED1} %
\definecolor{cbsix}  {HTML}{C85200} %
\definecolor{cbseven}{HTML}{898989} %
\definecolor{cbeight}{HTML}{A2C8EC} %
\definecolor{cbnine} {HTML}{FFBC79} %
\definecolor{cbten}  {HTML}{CFCFCF} %

\newcommand{\softfail}{\raisebox{0.5ex}{\texttildelow}}
\newcommand{\spfcode}[1]{\texttt{\small \color{cbone} #1}}

\begin{document}

\begin{acronym}[ECU]
	\acro{ASN}[ASN]{Autonomous System Number}
	\acro{SPF}[SPF]{Sender Policy Framework}
	\acro{SMTP}[SMTP]{Simple Mail Transfer Protocol}
	\acro{DNS}[DNS]{Domain Name System}
	\acro{DKIM}[DKIM]{DomainKeys Identified Mail}
	\acro{DMARC}[DMARC]{Domain-based Message Authentication, Reporting and Conformance}
	\acro{SMIME}[S/MIME]{Secure / Multipurpose Internet Mail Extensions}
	\acro{PGP}[PGP]{Pretty Good Privacy}
	\acro{MTA}[MTA]{Mail Transfer Agent}
	\acro{CDF}[CDF]{Cumulative Distribution Function}
	\acro{PoC}[PoC]{Proof of Concept}
	\acro{XSS}[XSS]{Cross-Site-Scripting}
    \acro{RRT}[RRT]{Resource Record Type}
\end{acronym}

\newcommand{\scannedDomains}{12823598}
\newcommand{\MXDomains}{9148550}
\newcommand{\SPFDomains}{7251736}
\newcommand{\SPFDomainsMX}{6868806}
\newcommand{\DMARCDomains}{1745566}
\newcommand{\NoMXSPFdenyallDomains}{202198}
\newcommand{\NoMXSPFsoftfailallDomains}{1143}
\newcommand{\IncludeDomains}{4856707}
\newcommand{\HugeDomains}{2517091}
\newcommand{\HugeIncludeDomains}{2507097}
\newcommand{\HugeNoIncludeDomains}{9994}
\newcommand{\ErrorDomains}{211018}
\newcommand{\MacroDomains}{194317}
\newcommand{\DeprecatedPTRDomains}{233167}
\newcommand{\DeprecatedSPFDomains}{107646}
\newcommand{\DNSRecordNotFound}{1179}
\newcommand{\SPFIncludeLoopOne}{13850}
\newcommand{\SPFSyntaxError}{38296}
\newcommand{\SPFTooManyDNSLookups}{49421}
\newcommand{\SPFTooManyVoidDNSLookups}{5308}
\newcommand{\SPFRedirectLoop}{58}
\newcommand{\SPFIncludeLoop}{19356}
\newcommand{\SPFRecordNotFound}{90697}
\newcommand{\SPFInvalidIP}{7882}
\newcommand{\NoDenyAll}{427767}
\newcommand{\ReportMechanism}{14}
\newcommand{\AllowAll}{10484}
\newcommand{\AllowInternet}{39}
\newcommand{\scannedDomainsM}{999997}
\newcommand{\SPFDomainsM}{602325}
\newcommand{\DMARCDomainsM}{226217}

\newcommand{\ScannedProviders}{56}

\newcommand{\SPFRescanSyntaxError}{36103}
\newcommand{\SPFRescanTooManyDNSLookups}{48630}
\newcommand{\SPFRescanTooManyVoidDNSLookups}{5127}
\newcommand{\SPFRescanRedirectLoop}{56}
\newcommand{\SPFRescanIncludeLoop}{18617}
\newcommand{\SPFRescanRecordNotFound}{88056}
\newcommand{\SPFRescanInvalidIP}{7498}
\newcommand{\SPFRescanResolvedErrors}{6931}
\newcommand{\NotificationsCount}{111951}
\newcommand{\RescanEmptyDomains}{1030}

\newcommand{\IncludesGeTen}{2408}
\newcommand{\IncludesGeTenDomains}{85915}
\newcommand{\IncludesBluehostCom}{68347}
\newcommand{\SyntaxErrorIP}{2946}
\newcommand{\SyntaxErrorIPvFour}{4216}
\newcommand{\SyntaxErrorIPvSix}{289}
\newcommand{\SyntaxErrorSiteVerification}{2699}
\newcommand{\SyntaxErrorMultipleSpf}{5847}
\newcommand{\SyntaxErrorWhitespace}{6344}
\newcommand{\RecordNotFoundSpfMissing}{48824}
\newcommand{\RecordNotFoundNotExisting}{36743}
\newcommand{\RecordNotFoundSpfMultiple}{2263}
\newcommand{\RecordNotFoundEmpty}{173}
\newcommand{\RecordNotFoundDNSError}{2691}
\newcommand{\RecordNotFoundOthers}{3}
\newcommand{\RecordNotFoundSpfMultipleCafeCom}{1711}

\newcommand{\IpFourDirect}{5583831}
\newcommand{\IpFourAll}{7007726}
\newcommand{\IpSixDirect}{197988}
\newcommand{\IpSixAll}{276288}
\newcommand{\IpFourDirectDomains}{2633991}
\newcommand{\IpFourAllDomains}{5741898}
\newcommand{\IpSixDirectDomains}{159500}
\newcommand{\IpSixAllDomains}{220386}
\newcommand{\TagsDirect}{38247985}
\newcommand{\TagsAll}{40313024}
\newcommand{\DomainsCidrZero}{54}
\newcommand{\DomainsCidrOne}{29}
\newcommand{\DomainsCidrTwo}{47}
\newcommand{\DomainsCidrThree}{16}
\newcommand{\DomainsCidrFour}{7}
\newcommand{\DomainsCidrFive}{6}
\newcommand{\DomainsCidrSix}{4}
\newcommand{\DomainsCidrSeven}{4}
\newcommand{\DomainsCidrEight}{2162}
\newcommand{\DomainsCidrNine}{23}
\newcommand{\DomainsCidrTen}{131}
\newcommand{\DomainsCidrEleven}{44}
\newcommand{\DomainsCidrTwelve}{313}
\newcommand{\DomainsCidrThirteen}{228}
\newcommand{\DomainsCidrFourteen}{1178}
\newcommand{\DomainsCidrFiveteen}{1145}
\newcommand{\DomainsCidrSixteen}{11126}
\newcommand{\IncludesCidrZero}{0}
\newcommand{\IncludesCidrOne}{2}
\newcommand{\IncludesCidrTwo}{10}
\newcommand{\IncludesCidrThree}{7}
\newcommand{\IncludesCidrFour}{3}
\newcommand{\IncludesCidrFive}{0}
\newcommand{\IncludesCidrSix}{0}
\newcommand{\IncludesCidrSeven}{0}
\newcommand{\IncludesCidrEight}{110}
\newcommand{\IncludesCidrNine}{3}
\newcommand{\IncludesCidrTen}{27}
\newcommand{\IncludesCidrEleven}{50}
\newcommand{\IncludesCidrTwelve}{137}
\newcommand{\IncludesCidrThirteen}{210}
\newcommand{\IncludesCidrFourteen}{5419}
\newcommand{\IncludesCidrFiveteen}{5389}
\newcommand{\IncludesCidrSixteen}{14243}

\newcommand{\NotificationAnswersThx}{300}
\newcommand{\NoMXDomains}{\FPeval{\result}{clip(\scannedDomains-\MXDomains)}\result\,\%\xspace}
\newcommand{\MXDomainsPerc}{\FPeval{\result}{round(clip(\MXDomains/\scannedDomains*100), 1)}\result\,\%\xspace}
\newcommand{\SPFDomainsPerc}{\FPeval{\result}{round(clip(\SPFDomains/\scannedDomains*100), 1)}\result\,\%\xspace}
\newcommand{\SPFDomainsMPerc}{\FPeval{\result}{round(clip(\SPFDomainsM/\scannedDomainsM*100), 1)}\result\,\%\xspace}
\newcommand{\SPFDomainsMXPerc}{\FPeval{\result}{round(clip(\SPFDomains/\MXDomains*100), 1)}\result\,\%\xspace}
\newcommand{\NoMXSPFDomainsPerc}{\FPeval{\result}{round((\SPFDomains-\SPFDomainsMX)/(\scannedDomains-\MXDomains)*100, 1)}\result\,\%\xspace}
\newcommand{\NoMXSPFdenyallDomainsPerc}{\FPeval{\result}{round(clip(\NoMXSPFdenyallDomains/(\scannedDomains-\MXDomains)*100), 1)}\result\,\%\xspace}
\newcommand{\NoMXSPFsoftfailallDomainsPerc}{\FPeval{\result}{round(clip(\NoMXSPFsoftfailallDomains/(\scannedDomains-\MXDomains)*100), 2)}\result\,\%\xspace}
\newcommand{\NoMXSPFrestrictallDomainsPerc}{\FPeval{\result}{round((\NoMXSPFsoftfailallDomains+NoMXSPFdenyallDomains)/(\SPFDomains-\SPFDomainsMX)*100, 1)}\result\,\%\xspace}
\newcommand{\DMARCAllDomainsPerc}{\FPeval{\result}{round(clip(\DMARCDomains/\scannedDomains*100), 1)}\result\,\%\xspace}
\newcommand{\DMARCAllDomainsMPerc}{\FPeval{\result}{round(clip(\DMARCDomainsM/\scannedDomainsM*100), 1)}\result\,\%\xspace}
\newcommand{\DMARCSPFDomainsPerc}{\FPeval{\result}{round(clip(\DMARCDomains/\SPFDomains*100), 1)}\result\,\%\xspace}
\newcommand{\IncludeDomainsPerc}{\FPeval{\result}{round(clip(\IncludeDomains/\SPFDomains*100), 1)}\result\,\%\xspace}
\newcommand{\ErrorDomainsPerc}{\FPeval{\result}{round(clip(\ErrorDomains/\SPFDomains*100), 1)}\result\,\%\xspace}
\newcommand{\HugeDomainsPerc}{\FPeval{\result}{round(clip(\HugeDomains/\SPFDomains*100), 1)}\result\,\%\xspace}
\newcommand{\MacroDomainsPerc}{\FPeval{\result}{round(clip(\MacroDomains/\SPFDomains*100), 1)}\result\,\%\xspace}
\newcommand{\SPFIncludeLoopOnePerc}{\FPeval{\result}{round(clip(\SPFIncludeLoopOne/\SPFIncludeLoop*100), 1)}\result\,\%\xspace}
\newcommand{\SPFSyntaxErrorPerc}{\FPeval{\result}{round(clip(\SPFSyntaxError/\ErrorDomains*100), 2)}\result\,\%\xspace}
\newcommand{\SPFTooManyDNSLookupsPerc}{\FPeval{\result}{round(clip(\SPFTooManyDNSLookups/\ErrorDomains*100), 2)}\result\,\%\xspace}
\newcommand{\SPFTooManyVoidDNSLookupsPerc}{\FPeval{\result}{round(clip(\SPFTooManyVoidDNSLookups/\ErrorDomains*100), 2)}\result\,\%\xspace}
\newcommand{\SPFRedirectLoopPerc}{\FPeval{\result}{round(clip(\SPFRedirectLoop/\ErrorDomains*100), 2)}\result\,\%\xspace}
\newcommand{\SPFRecordNotFoundPerc}{\FPeval{\result}{round(clip(\SPFRecordNotFound/\ErrorDomains*100), 2)}\result\,\%\xspace}
\newcommand{\SPFIncludeLoopPerc}{\FPeval{\result}{round(clip(\SPFIncludeLoop/\ErrorDomains*100), 2)}\result\,\%\xspace}
\newcommand{\SPFInvalidIPPerc}{\FPeval{\result}{round(clip(\SPFInvalidIP/\ErrorDomains*100), 2)}\result\,\%\xspace}
\newcommand{\RecordNotFoundSpfMissingPerc}{\FPeval{\result}{round(clip(\RecordNotFoundSpfMissing/\SPFRecordNotFound*100), 1)}\result\,\%\xspace}
\newcommand{\RecordNotFoundNotExistingPerc}{\FPeval{\result}{round(clip(\RecordNotFoundNotExisting/\SPFRecordNotFound*100), 1)}\result\,\%\xspace}
\newcommand{\RecordNotFoundSpfMultiplePerc}{\FPeval{\result}{round(clip(\RecordNotFoundSpfMultiple/\SPFRecordNotFound*100), 1)}\result\,\%\xspace}
\newcommand{\RecordNotFoundEmptyPerc}{\FPeval{\result}{round(clip(\RecordNotFoundEmpty/\RecordNotFoundEmpty*100), 1)}\result\,\%\xspace}
\newcommand{\RecordNotFoundDNSErrorPerc}{\FPeval{\result}{round(clip(\RecordNotFoundDNSError/\SPFRecordNotFound*100), 1)}\result\,\%\xspace}
\newcommand{\RecordNotFoundSpfMultipleCafeComPerc}{\FPeval{\result}{round(clip(\RecordNotFoundSpfMultipleCafeCom/\RecordNotFoundSpfMultiple*100), 1)}\result\,\%\xspace}
\newcommand{\SyntaxErrorIPPerc}{\FPeval{\result}{round(clip(\SyntaxErrorIP/\SPFSyntaxError*100), 1)}\result\,\%\xspace}
\newcommand{\SyntaxErrorIPvFourPerc}{\FPeval{\result}{round(clip(\SyntaxErrorIPvFour/\SPFSyntaxError*100), 1)}\result\,\%\xspace}
\newcommand{\SyntaxErrorIPvSixPerc}{\FPeval{\result}{round(clip(\SyntaxErrorIPvSix/\SPFSyntaxError*100), 1)}\result\,\%\xspace}
\newcommand{\SyntaxErrorSiteVerificationPerc}{\FPeval{\result}{round(clip(\SyntaxErrorSiteVerification/\SPFSyntaxError*100), 1)}\result\,\%\xspace}
\newcommand{\SyntaxErrorMultipleSpfPerc}{\FPeval{\result}{round(clip(\SyntaxErrorMultipleSpf/\SPFSyntaxError*100), 1)}\result\,\%\xspace}
\newcommand{\SyntaxErrorWhitespacePerc}{\FPeval{\result}{round(clip(\SyntaxErrorWhitespace/\SPFSyntaxError*100), 1)}\result\,\%\xspace}
\newcommand{\NotificationSuccess}{\FPeval{\result}{round(clip(\SPFRescanResolvedErrors/\ErrorDomains*100), 1)}\result\,\%\xspace}
\newcommand{\NoDenyAllPerc}{\FPeval{\result}{round(clip(\NoDenyAll/\SPFDomains*100), 1)}\result\,\%\xspace}
\newcommand{\IncludesBluehostComPerc}{\FPeval{\result}{round(clip(\IncludesBluehostCom/\IncludesGeTenDomains*100), 1)}\result\,\%\xspace}
\newcommand{\IpSixDomainsPerc}{\FPeval{\result}{round(clip(\IpSixDirect/\TagsDirect*100), 1)}\result\,\%\xspace}

\title[Lazy Gatekeepers]%
      {Lazy Gatekeepers: A Large-Scale Study\\ on SPF~Configuration  in the Wild}

\author{Stefan Czybik}
\affiliation{%
 \institution{Technische Universität Berlin}
 \department{Machine Learning and Security}
 \city{Berlin}
 \country{Germany}
}

\author{Micha Horlboge}
\affiliation{%
 \institution{Technische Universität Berlin}
 \department{Machine Learning and Security}
 \city{Berlin}
 \country{Germany}
}

\author{Konrad Rieck}
\affiliation{%
 \institution{Technische Universität Berlin}
 \department{Machine Learning and Security}
 \city{Berlin}
 \country{Germany}
}

\renewcommand{\shortauthors}{Stefan Czybik, Micha Horlboge, and Konrad Rieck}

\begin{abstract}
The \acl{SPF} (SPF) is a basic mechanism for authorizing the use of domains in email.
In combination with other mechanisms, it serves as a cornerstone for protecting users from forged senders.
In this paper, we investigate the configuration of SPF across the Internet. To this end, we analyze SPF records from 12~million domains in the wild. %
Our analysis shows a growing adoption, with \SPFDomainsPerc of the domains providing SPF records. 
However, we also uncover notable security issues: First, \ErrorDomainsPerc of the SPF records have errors, undefined content or ineffective rules, undermining the intended protection.
Second, we observe a large number of very lax configurations. For example, \HugeDomainsPerc of the domains allow emails to be sent from over \numprint{100000}~IP~addresses. 
We explore the reasons for these loose policies and demonstrate that they facilitate email forgery.
As a remedy, we derive recommendations for an adequate configuration and notify all operators of domains with misconfigured SPF records.
\end{abstract}


\keywords{Email, SPF, Authorization, Email Forgery}

\maketitle
\acresetall
\section{Introduction}

Email still represents the prime form of communication on the Internet today.
Despite several weaknesses of the protocol, billions of users regularly use email messages for business and personal exchange~\citep{emailUsersWorld}.
Due to its popularity, email is a constant magnet for cybercrime, serving as a vehicle for transporting unsolicited, fraudulent and malicious content, which ranges from spam and phishing attempts to targeted attacks and malware distribution~\mbox{\citep[e.g.,][]{ SimZanThoBur+20, KanKreLevEnr+08, 
HuWan18, GasUllStrRie+18}}.
These activities benefit from the lack of security mechanisms in the original protocols that cannot establish the authenticity of senders and content by itself.

Several extensions have been proposed over the last years to counter the misuse of email, including security mechanisms for the transport layer \citep{rfc3207, rfc2595}, email headers \citep{rfc6376}, and message data \citep{rfc8551, rfc3156}.
One of the oldest mechanisms to mitigate the spoofing of email senders is the \emph{Sender Policy Framework} (SPF)~\citep{rfc7208}.
Instead of retrieving emails from any network host, the receiving server can request an SPF record from the sender's domain and check whether the connecting IP address is authorized to send emails.
In concert with other mechanisms, such as DKIM~\citep{rfc6376} and DMARC~\citep{rfc7489}, SPF forms one central pillar for mitigating forged emails.

Despite this important role, however, the configuration of SPF in the wild and its weak spots are still an open field of research.
The study by \citet{Gojmerac2015} from 2014 indicates a moderate adoption of the mechanism and a tendency towards coarse authorization.
Further studies in the following years show an increasing number of domains using SPF.
In this work, we expand this view on SPF and present a detailed analysis of its configuration on the Internet.
In particular, we use the \mbox{Tranco list \citep{TrancoList}} to collect SPF records from 12 million domains over a period of 5 months.
Based on this collection, we analyze the adoption, validity and permissiveness of SPF policies to learn how servers use this mechanism.

Our study reveals a growing adoption of SPF in practice.
While \citeauthor{Wang2022} report in 2022 that 54.1\,\% of the domains contain valid SPF records, we observe an adoption of \SPFDomainsMPerc for the top 1 million and \SPFDomainsPerc for all 12 millions domains in our study.
Unfortunately, we also uncover persisting security issues: First, \ErrorDomainsPerc of the SPF records suffer from errors, undefined content, or ineffective rules, undermining the intended protection.
Second, we observe a large number of \emph{very} lax configurations.
For example, \HugeDomainsPerc of the domains allow emails to be sent from over \numprint{100000} IP addresses.
We demonstrate in a case study that these coarse configurations give rise to spoofing email senders and thus unnecessarily weaken the protection of SPF in the Internet.

To mitigate this situation, we investigate the reasons for the lax policies and derive guidelines for a more restrictive configuration. Moreover, we have launched a notification campaign for all SPF records with invalid policies. In total, we have contacted \numprint{\NotificationsCount} operators by email and informed them about incorrect or insufficient configurations. Feedback on these reports has been positive, and several operators promised to fix the reported problems.
Two weeks after our notification, a scan of the domains shows that \numprint{\SPFRescanResolvedErrors} (\NotificationSuccess) of the entries have already been corrected, and we expect further improvement over the next months. 

\smallskip
\textbf{Roadmap.} We review the background of SPF in \autoref{sec:background} and discuss related studies in \autoref{sec:previous_work}.
Afterward, we describe the methodology of our study in \autoref{sec:methodology}.
Our findings are presented in Sections \ref{sec:errors} and \ref{sec:includes}, where we first investigate invalid configurations and then explore the coarse use of SPF authorization.
Our guidelines are presented in \autoref{sec:recommendations}, before we conclude in \autoref{sec:conclusions}.

\section{Background}
\label{sec:background}

The sending and receiving of email is realized on top of the classic \ac{SMTP}~\citep{rfc821}.
Standardized in 1982, this protocol has been designed without built-in mechanisms to ensure the confidentiality of transmitted messages or to verify the authenticity of senders.
As the importance and ubiquity of email has grown over time, the need for enhanced security measures has become increasingly apparent.

One significant security concern is the propagation of emails with forged sender addresses, for example, as part of spam and phishing campaigns. 
These forged emails exploit the lack of authenticity in \ac{SMTP} and are a notorious threat to users.
In response to this security gap, the \ac{SPF}~\citep{rfc4408} was introduced in 2003 as a standard to define approved sending servers of emails for a specific domain.
To this end, a domain owner can configure a \ac{DNS} record, which specifies a list of authenticated hostnames or IP addresses that are permitted to send emails on behalf of the domain.

While the introduction of \ac{SPF} appears reasonable at first glance, it does face certain limitations.
Primarily, SPF only addresses the authenticity problem by extending it to hostnames and IP addresses.
This means that users have to trust their network provider or managed email service to accurately handle this aspect of email security.
Furthermore, SPF introduces new problems when it comes to email redirection, which becomes problematic for mailing lists.

As another protocol to improve the authenticity of emails, \ac{DKIM} requires the sending server to add a cryptographic signature to all outgoing emails.
These signatures can then be verified by the receiving mail server, providing an additional layer of authenticity.
On top of DKIM and SPF, \ac{DMARC}~\citep{rfc7489} adds a descriptive record to the DNS.
This entry describes the behavior that a receiving mail server should adopt when an email is received and there are issues with SPF or DKIM authentication.

Note that \ac{SPF} as well as \ac{DKIM}, and \ac{DMARC} are not able to provide reliable confidentiality, integrity, and authenticity for end-to-end communications like S/MIME~\citep{rfc8551} and OpenPGP~\citep{rfc3156}.
These mechanisms, however, are not widely adopted yet and suffer from their own problems~\citep{Poddebniak2018}. Unlike SPF, which is implemented in the application layer, these mechanisms are implemented on top of email and do not affect email servers. 
Consequently, large email providers, such as Google and Microsoft, recommend and enforce the use of \ac{SPF} as a basic element of email security. 

\subsection{A Primer on SPF}

The security mechanism \ac{SPF} operates through \ac{DNS} records that store a configuration of permitted IP addresses, networks or hostnames. This configuration is controlled by the domain owner and is publicly accessible.
When a server receives an incoming email, it can perform a DNS lookup to retrieve the corresponding \ac{SPF} record associated with the sender's domain.
While processing the configuration, the server validates whether the email originates from an authorized source. 
In the following, we use the term \textit{SPF record} to refer to the string in a DNS request of type \emph{TXT} that starts with \spfcode{v=spf1} and defines the configuration. The deprecated DNS type \emph{SPF} is not considered in this work.

Technically, an SPF record is composed of different policy terms.
These terms are either directives containing \emph{mechanisms} with qualifiers, or \emph{modifiers}.
While modifiers provide additional information for the configuration of the policy, a mechanism defines a way to determine allowed IP addresses, networks or hostnames.
Once there is a match between the sending host and a mechanism directive, the processing of the SPF record ends and the mechanism's qualifier is returned as the result of the authorization. 

\paragraph{Mechanisms.}
We first take a look at the different mechanisms and their qualifiers.

\begin{description}
\setlength{\itemsep}{2pt}
\item[\spfcode{a}] \hfill \\
This mechanism matches if the sending IP addresses match the specified A or AAAA DNS records.

\item[\spfcode{mx}]\hfill \\
If an email originates from any of the hostnames or IP addresses specified in an MX DNS record, this mechanism matches.

\item[\spfcode{ip4}, \spfcode{ip6}]\hfill \\
It is also possible to set allowed IP addresses.
If the sender IP is listed here, this mechanism matches.

\item[\spfcode{all}]\hfill \\
As the name says, this mechanism matches all sender IPs.
Everything after this term is ignored.

\item[\spfcode{exists}]\hfill \\
This mechanism can check if a specific domain or hostname exists in the DNS.
If the hostname exists, this mechanism matches.

\item[\spfcode{include}]\hfill \\
The \spfcode{include} mechanism allows a domain to include another domain's permitted sender IPs from its SPF record.
This is useful to cross administrative borders at email delivery.
The receiving server evaluates the content of the included SPF record as usual, but this mechanism only matches if the sender IP is explicitly allowed by it.
Otherwise, and also in case of an error, the processing of the including record continues.
Therefore, it is not possible to deny any or all IP addresses with the \spfcode{include} mechanism.

\item[\spfcode{ptr}]\hfill \\
The last one, the \spfcode{ptr} mechanism, checks if a reverse DNS entry for the sending IP address exists.
This mechanism matches if the IP addresses of the sending host and of the domain name retrieved by the reverse lookup are equal.
Since this is a slow mechanism that causes a high DNS load, using this mechanism is generally not recommended.

\end{description}

Except for \spfcode{all}, the mechanisms can be specified by arguments. If no argument is given, the domain or IP address to be checked is used.
The \spfcode{a}, \spfcode{mx} and both \spfcode{ip} mechanisms additionally allow specifying a CIDR prefix length to  specify a complete network.
If no CIDR prefix length is given, it will refer to a single host.

A qualifier can be placed in front of each mechanism to define the outcome in case the IP address of the sending email server matches.
If a mechanism is specified without a qualifier, \spfcode{pass} is implied.

\begin{description}
\setlength{\itemsep}{2pt}
\item[\spfcode{+} \hspace{5pt}{\mdseries (pass)}]\hfill \\
{
The email server is authorized to send emails for the domain.
}

\item[\spfcode{-} \hspace{5pt}{\mdseries (fail)}]\hfill \\
{
The email server is explicitly not authorized to send emails for the domain.  
}

\item[\spfcode{?} \hspace{5pt}{\mdseries (neutral)}]\hfill \\
{
There is no assertion about the email server.
}

\item[\spfcode{\softfail} \hspace{4pt}{\mdseries (softfail)}]\hfill \\
{
The email server is neither explicitly denied nor allowed to send emails for the domain. It is not authorized, but not strong enough to create a strict policy.
}
\end{description}

If the evaluation has found a match between the sending IP address and a mechanism, the qualifier is returned as a result and gives information about the authenticity of the sender.
Note that the default result for \ac{SPF} is not \spfcode{fail}.
If there is no explicit \spfcode{fail} or \spfcode{softfail} qualifier for the \spfcode{all} mechanism, the SPF result for all hosts without another match is always the default value \spfcode{pass}.
If no mechanism matches, for example, because the IP address is not listed as approved sender and there is no \spfcode{all} mechanism set, the result is \spfcode{neutral}.

\paragraph{Modifiers.}
In addition to the mechanisms, there are also modifiers, of which for our work only the \spfcode{redirect} modifier is relevant. 
This modifier allows a domain to delegate its SPF record to another domain.
Like the \spfcode{include} mechanism, this can be used to cross administrative borders, but in contrast to that mechanism, the complete evaluation process is performed on the redirected domain. Any statements after a \spfcode{redirect} modifier are ignored. \\

Additionally, the evaluation of an SPF record at the receiving site can provide further return values.
In particular, \spfcode{none} is returned when there is no valid domain from the SMTP session or no SPF record.
In the event of a transient error like a DNS error, a \spfcode{temperror} is raised.
If a DNS error is permanent, such as NXDOMAIN, a \spfcode{permerror} is returned.
The \spfcode{permerror} is further used when the SPF record can not be evaluated correctly.
In \autoref{sec:errors}, we investigate the occurrence of these errors in the wild.

\paragraph{Example.}
Let us investigate the following SPF record:
\begin{center}
\color{cbone}
\begin{verbatim}
 v=spf1 +mx a:puffin.example.com/28 -all
\end{verbatim}
\end{center}
In this example, we have multiple mechanisms: \spfcode{mx}, \spfcode{a} and \spfcode{all}.
The term \spfcode{+mx} with the explicit \spfcode{pass} qualifier specifies that the domain's MX servers are authorized to send mails.
The next directive specifies an IP address range, namely the IP address of \spfcode{puffin.example.com} with a /28 CIDR notation.
As no explicit qualifier is given, the default \spfcode{pass} is used, and all addresses in this range are authorized.
The last directive, \spfcode{-all}, enforces to reject emails from all other sources.

\section{Previous Studies}
\label{sec:previous_work}
Since email security is an important topic, we are not the first to measure the prevalence of sender authentication mechanisms.
Over the last decade, the adoption of these techniques and possible vulnerabilities have been studied several times.
In the following, we briefly review this work.

\medskip
\emph{Studies from 2014 to 2018.}
In 2014, \citet{Gojmerac2015} scanned the top 1 million of the Alexa ranking for DNS entries such as SPF and DMARC.
Only about 37\,\% of the domains provided an SPF configuration at that time.
In addition, \citeauthor{Gojmerac2015} found several common syntactic errors in SPF records in their study, such as missing values for matching mechanisms like \spfcode{ip4}, but did not quantify them further.
In 2015, \citet{Durumeric_2015} gave a broad overview of the adoption rates of security extensions for \ac{SMTP}.
Besides protocols like STARTTLS and \ac{DKIM}, they also investigated \ac{SPF} for the Alexa top 1 million list but ignored sites without an MX record.
They found that 47\,\% of the domains had published an SPF policy, indicating a growing adoption.

In the same year, \citet{Foster_2015} evaluated the security provided against network attacks by such extensions from a theoretical and practical view.
As part of their study, they also scanned the Alexa list and, additionally, the top million mail domains from a leaked set of user data from Adobe.
The result here was that 42.26\,\% of the Alexa domains and 43.60\,\% of the Adobe domains were using \ac{SPF}.
A few years later, \citet{HuWan18} investigated how email providers handle spoofed emails and if such could reach the inbox of the users.
In this context, they searched the top 1 million domains from the Alexa ranking for \ac{SPF} records and reported a slightly increased adoption rate of 44.9\,\%. 

\medskip
\emph{Studies from 2020 to 2023.}
\citet{Tatang2021} measured the adoption rate of SPF, DKIM and DMARC and analyzed the relationship between different domains through included SPF entries as well as the domains and the autonomous systems belonging to the allowed IP addresses.
Therefore, they scanned in 2020 over 2 million domains from different top lists, of which 50.7\,\% had published SPF records.
Moreover, they reported 13\,\% invalid entries and, as the most common error, too many DNS lookups. %
The authors also mentioned that many records used different includes, and that sometimes large IP subnets are trusted.

In the same year, \mbox{\citet{Kahraman2020}} analyzed the usage of \ac{SPF} on a dataset of about 168 million domains. In this very large dataset, 25\,\% of the domains had \ac{SPF} configured and were further analyzed in terms of the used mechanisms and syntactic as well as DNS lookup limit errors.
\citet{Trost2020} crawled, also in 2020, about 8.3 million domains from different top lists for SPF records to analyze trust relationships.
The analysis showed that some domains allowed very large networks to send emails on behalf of them, raising concerns about possible attack vectors.
The measurements in these three papers are close to ours, yet we provide a detailed analysis of the \ac{SPF} records themselves, which allows us to characterize the security risks and notify the affected operators.

Two years later, \citet{Wang2022} measured the deployment of \ac{DKIM} and issues of the management.
In their work, they also reported an \ac{SPF} adoption rate of 54,1\,\% in the Alexa top 1 million domains.
This result continued the trend from previous work, that has shown an increasing number of domains with such a policy.

\paragraph{Attacks on SPF}
\citet{Deccio2021} analyzed how email servers process and validate \ac{SPF} entries. They observed that several servers ignore syntax errors and ambiguities of the specification, which could lead to various forms of attacks.
In a similar vein, \citet{Shen2021} investigated several security protocols, including SPF and developed attacks for the authentication by systematically exploiting details in the standards that are often implemented inconsistently.
They proposed more accurate protocol descriptions to eliminate the ambiguous definitions, which in the end could also decrease the number of errors in DNS records.
Another attack vector was described by \citet{Liu2023}:
In their work, they investigated different types of email forwarding and how these change header fields.
In the end, the implementation of some forwarding techniques enabled the authors to circumvent methods like \ac{SPF} and to send spoofed emails without detection.

Finally, the implementation of sender validation libraries itself can be a point of attack.
\citet{Bennett2022} demonstrated this using \textit{libSPF2} as an example.
They found multiple bugs in it and developed a technique to detect vulnerable servers remotely, revealing the widespread use of this library version.

\paragraph{Difference to our study.}
Our study continues the line of previous research and extends it with additional perspectives:
We base our study on a larger dataset than most previous studies, except for the work by \citet{Kahraman2020}.
This gives us a broader picture of the use of \ac{SPF} in the wild.
As a result, we are able to perform a detailed analysis of the flaws and weaknesses in SPF configurations, showing where and why authentication fails.
This combination of a large dataset and detailed analysis provides valuable insights into common problems when applying the SPF framework.
Moreover, we conduct a case study demonstrating that overly coarse authorization policies weaken the security mechanism and make it easier to forge emails with spoofed senders.

\section{Methodology}
\label{sec:methodology}

Next, we introduce our methodology for investigating SPF records, their errors and potential threats.
The goal of our study is threefold:
First, with a large-scale measurement, we want to determine the prevalence of SPF across a wide range of domains.
Second, we aim to shed light on how often and why SPF entries are flawed and thus only provide inadequate protection.
Finally, we want to assess the occurrence and impact of overly coarse authorizations in SPF configurations.

\subsection{Measuring SPF in the Wild}
\label{sec:measuring-in-the-wild}

From a technical perspective, we have two options to measure the configuration of \ac{SPF}: 
As the first strategy, we can collect a representative set of emails and extract all sender and recipient email addresses.
From these addresses, we could generate a list of domains of email providers and examine their \ac{SPF} records, similar to the study of \mbox{\citet{Durumeric_2015}}. 
Second, we can use a list of domains and retrieve all available \ac{SPF} records from them, even if they are not intended to ever be used to send emails.
While the first strategy helps to understand how \ac{SPF} is used relative to the distribution of email providers, the latter one provides a less biased view of \ac{SPF} configuration in the wild.
Consequently, we pursue this strategy for our large-scale study.

\paragraph{Data source.}
We use the Tranco list of domains~\citep{TrancoList} for our measurements.
This list is a research ranking of well-known and frequently used websites.
We use the full lists of the first of the months from January until May 2023 and merge them to get a bigger amount of domains.

\paragraph{Crawler.}
We develop a crawler for collecting and parsing SPF records using the \emph{checkdmarc library}\footnote{Available at \url{https://github.com/domainaware/checkdmarc}}.
The crawler retrieves the SPF record for a given domain using the function \spfcode{query\_spf\_record()}.
This function sends DNS requests of type TXT and SPF, but only returns the first SPF record from the type TXT request.
The record is then parsed using \spfcode{parse\_spf\_record()}.
To analyze different weaknesses, flaws and misconfigurations, we modify the library.
Our modified version returns all necessary values, such as the number of DNS lookups, permitted IP addresses and a parsed version of the SPF record.
Warnings and errors in the SPF syntax are reported, and our modified version continues with the parsing afterward.
Due to the scale of our study, we implement a cache to reduce the DNS load by not sending the same request twice.
If an SPF record already exists in the database, the cached object is used instead of requesting and analyzing it again.
This reduces the load from include mechanisms of large providers significant.
Only for the first domain the include mechanism is processed, all others hit the cache. 
Moreover, we distribute and rate limit the DNS requests across 150 servers.
The same procedure is applied for \ac{DMARC} using \spfcode{query\_dmarc\_record()} and \spfcode{parse\_dmarc\_record()}.
In the end, we collect the following information per domain:
\begin{itemize}
	\item SPF record 
	\item DMARC record
	\item MX record 
\end{itemize}
Note that this information is publicly available and therefore no confidential or private data is collected in our study, see also Appendix~\ref{app:ethics}.

We then analyze the collected records by checking for errors and misconfigurations.
Moreover, we evaluate the matching mechanisms of \ac{SPF} and investigate the resulting authorization policy.
For example, we determine the amount and type of authorized senders by recursively analyzing the \spfcode{include} mechanism.

\paragraph{Measurement focus.}
The main focus of our study is to understand the configuration of SPF entries and the role of authorized hosts in the underlying policies. 
By analyzing the collected data, we can examine these properties in detail. However, there are also limitations resulting from our study design:
First, we can analyze all SPF mechanisms except for \spfcode{exist}.
This can only be done with the first measurement strategy and a dataset of representative emails. 
Second, we restrict our study to IPv4 hosts. 
\citeauthor{Durumeric_2015} report that only 1.13\,\% of the mechanisms in SPF are \spfcode{ip6} terms. 
In our scan, we find an even lower adoption rate.
Only \IpSixDomainsPerc of the domains use IPv6 directly, which is why we refrain from a detailed analysis.

\section{SPF Adoption and Errors}
\label{sec:errors}

We begin our examination of the collected data by first analyzing the adoption of SPF in the wild and comparing it to previous work. We then proceed with a detailed analysis of the uncovered errors and misconfigurations, expanding the scope of previous studies.

\subsection{SPF Usage}

In total, we have scanned \numprint{\scannedDomains} domains for this study.
While this expanded scan provides insights on the general configuration of SPF, it is not directly comparable with previous studies that have considered smaller sets from the top 1 million domains of the Alexa and other rankings.
However, in our measurement, the result for the top 1 million domains is included. Thereby, it is comparable in terms of size and the fact that the domains are ranked.
Therefore, we first focus on the top 1 million domains of the \mbox{Tranco list\footnote{Available at \url{https://tranco-list.eu/list/K2XYW}.} \citep{TrancoList}} generated on 01 May 2023. %

\begin{figure}[b]
	\centerline{\input{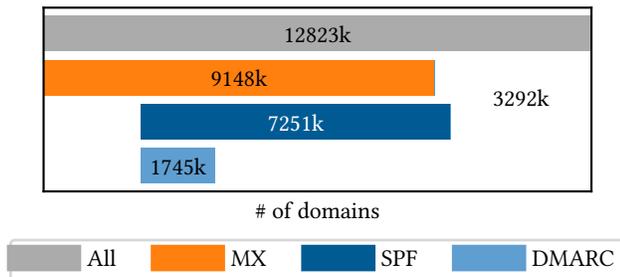}}
	\vspace{-6pt}
	\caption{Implementation of email and security mechanisms and their overlaps.}
	\label{fig:domains_overview}
\end{figure}

Using this focus, we observe that the usage of SPF per domain has grown to \SPFDomainsMPerc of all scanned domains and \SPFDomainsMXPerc for domains with MX record.
A detailed comparison of our results with past measurements is shown in \autoref{tab:usage}.
We find that domains within the first 1 million use SPF and DMARC more frequently.
But also for the complete 12 million domains, a clear increase in SPF usage to \SPFDomainsPerc can be observed from our scan. Every second domain in our measurement is now employing this security mechanism.
Additionally,  \autoref{fig:domains_overview} provides an overview of all scanned domains and their adoption of SPF and DMARC.

\begin{table}[tbp]
\begin{threeparttable}
	\centering \small
	\caption{SPF and DMARC usage in the wild.}
	\label{tab:usage}
	\begin{tabular}{lclrrr}
		\toprule
		\textbf{Study}            & \textbf{Year} & \textbf{List} & \textbf{Size} & \textbf{SPF}     & \textbf{DM.}           \\
		\midrule
        \citet{Gojmerac2015}    & 2014          & Alexa         & 1M               & 36.7\,\%        & 0.5\,\%             \\
		\citet{Foster_2015}     & 2015          & Alexa         & 1M               & 42.2\,\%        & 1.0\,\%             \\
		\citet{Foster_2015}     & 2015          & Adobe         & 1M               & 43.6\,\%        & 0.9\,\%             \\
		\citet{Durumeric_2015}\tnote{1}  & 2015          & Alexa         & 1M               & 47.0\,\%           & 1.1\,\%              \\
        \citet{HuWan18}          & 2018          & Alexa         & 1M               & 49.2\,\%         & 5.1\,\%              \\
        \citet{Kahraman2020}    & 2020          & Alexa          & 1M & 73.6\,\% & --- \\
        \citet{Wang2022} & 2022 & Alexa & 1M & 54.1\,\% & 11.9\,\% \\
		Our study                     & 2023          & Tranco        & 1M               & \SPFDomainsMPerc & \DMARCAllDomainsMPerc \\
        \midrule
        \citet{Tatang2021}      & 2020          & Other\tnote{2} & 2M & 50.7\,\%  & 11.5\,\% \\
        \citet{Kahraman2020}    & 2020          & None          & 168M & 25.0\,\% & --- \\
		Our study                     & 2023          & Tranco        & 12M              & \SPFDomainsPerc  & \DMARCAllDomainsPerc  \\
		\bottomrule
	\end{tabular}
    \begin{tablenotes}
        \item[1] Only domains with MX record are considered in the evaluation
        \item[2] Union of Alexa, Majestic and Tranco top 1M lists
    \end{tablenotes}
 \end{threeparttable}
\end{table}

We also observe an interesting phenomenon: \NoMXSPFDomainsPerc of the domains without an MX record return an \ac{SPF} record. At first glance, this may seem counterintuitive, since these domains specify which senders are authorized through SPF but cannot receive email themselves. In several cases, these domains are not intended to send or receive email, and so the SPF record is used to deny sending email in general. We find that \NoMXSPFrestrictallDomainsPerc of the domains without an MX but an SPF record have SPF configurations containing \spfcode{v=spf1 -all} (\numprint{\NoMXSPFdenyallDomains}) or \spfcode{v=spf1 \softfail all} (\numprint{\NoMXSPFsoftfailallDomains}). However, the remaining half of these domains are likely misconfigured because they specify a sending policy but cannot receive bounces or other error messages from the transport, making them unsuitable for reliable email communication.

\subsection{DMARC}
In addition to SPF, we have also scanned for \ac{DMARC} records using the \emph{checkdmarc library}.
We have done this to measure the increment from previous studies on email security.
As shown in \autoref{tab:usage}, \acp{DMARC} started with a low value of about 1\,\% in 2015 and is now at \DMARCAllDomainsMPerc for the top 1 million domains and \DMARCAllDomainsPerc for all domains. 
The increasing usage of \ac{DMARC} is likely due to the recommendations of large email providers\footnote{\url{https://support.google.com/a/answer/2466580?hl=en}}.
\citeauthor{Durumeric_2015} already mentioned, that major email providers, such as Google and Microsoft, heavily skew the apparent adoption of security mechanisms.

\subsection{SPF Errors}

\begin{figure}[htbp]
	\centerline{\input{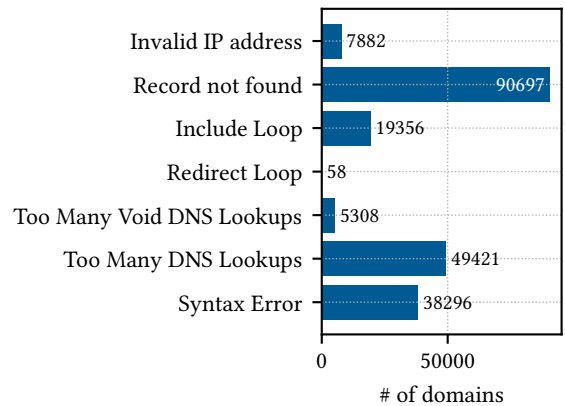}}
	\vspace{-6pt}
	\caption{Appearance of different error types.}
	\label{fig:spf_errors}
\end{figure}

In our analysis of SPF, we observe a variety of errors in \ErrorDomainsPerc (\numprint{\ErrorDomains}) of the domains, some of which are trivial typos while others are rather subtle misconfigurations.
We hence explore these errors in more detail and count all issues as errors that affect the correct functionality of SPF.
This includes all records that might result in a \spfcode{permerror}.
\autoref{fig:spf_errors} provides a general overview of all types of errors found.

Note that during our scan, we received \numprint{\DNSRecordNotFound} DNS errors.
This means that a domain that was supposed to be resolved when parsing the SPF record was not resolvable at that time. Since this may change on subsequent scans, we exclude these errors from the following analysis.

\paragraph{Record not found}
First, we consider record-not-found errors indicating that no SPF record was found for a given domain name. These errors are the most common in our study with \SPFRecordNotFoundPerc.
They can be caused by either the \spfcode{include} mechanism or the \spfcode{redirect} modifier, since the SPF record of another domain must be parsed.

\begin{figure}[htbp]
	\centerline{\input{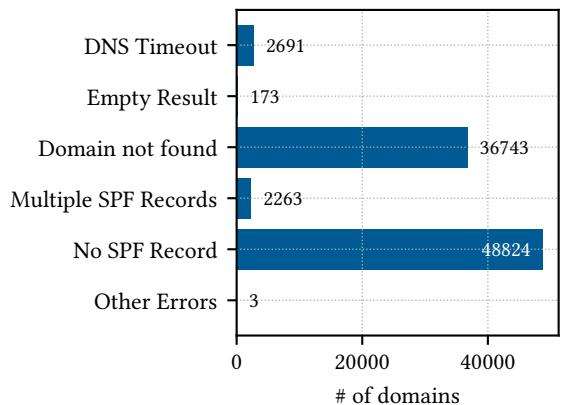}}
	\vspace{-6pt}
	\caption{Distribution of record-not-found errors.}
	\label{fig:record_errors}
\end{figure}

If we look at this error in detail, we see in \autoref{fig:record_errors} that there are different causes.
The most common cause with \RecordNotFoundSpfMissingPerc (\numprint{\RecordNotFoundSpfMissing}) of this error type is, that the requested domains has no SPF record. 
In contrast, there are \RecordNotFoundSpfMultiplePerc (\numprint{\RecordNotFoundSpfMultiple}) that provide more than one SPF record, which is no valid SPF record by the specification.
An interesting finding here is that \RecordNotFoundSpfMultipleCafeComPerc (\numprint{\RecordNotFoundSpfMultipleCafeCom}) of these errors are due to an \spfcode{include} of the provider \textit{cafe24.com}, which is a hosting provider for business customers.
Another common record-not-found error is that the requested domain is not found (NXDOMAIN), as it happens \numprint{\RecordNotFoundNotExisting} (\RecordNotFoundNotExistingPerc) times.
This error could become critical if the domain is not registered and is taken over by an attacker.
Other DNS related errors like a timeout, what is a \spfcode{temperror} or an empty result are less common.
The three other errors are one each of a DNS label is > 63 octets long, a DNS name is > 255 octets long, and one utf-8 decode error. 

\paragraph{Too many DNS lookups}
To prevent denial-of-service attacks, the number of DNS lookups that an SPF record may trigger is limited to 10 requests.
It is the second commonest error with \SPFTooManyDNSLookupsPerc.
As the specification is not totally clear here, we need to discuss this error in detail.
RFC7208 says:
\begin{quote}\color{cbone}
The following terms cause DNS queries: the "include", "a", "mx", "ptr", and "exists" mechanisms, and the "redirect" modifier.
SPF implementations MUST limit the total number of those terms to 10 during SPF evaluation, to avoid unreasonable load on the DNS.
\end{quote}

The problem here is that for the \spfcode{include} mechanism, there is no further description of how recursive DNS requests should be handled.
As for \spfcode{mx} and \spfcode{ptr} mechanisms they are within the overall limit of 10, we assume this holds for the \spfcode{include} mechanism too.
In the \emph{checkdmarc} library, this is implemented by counting the mechanism-related lookups during recursion.
Another important fact is that this error does not have to lead directly to a \spfcode{permerror} in the SPF check. The SPF check can be successful if a result is returned within the first 10 lookups.

\begin{figure}[htbp]
	\centerline{\input{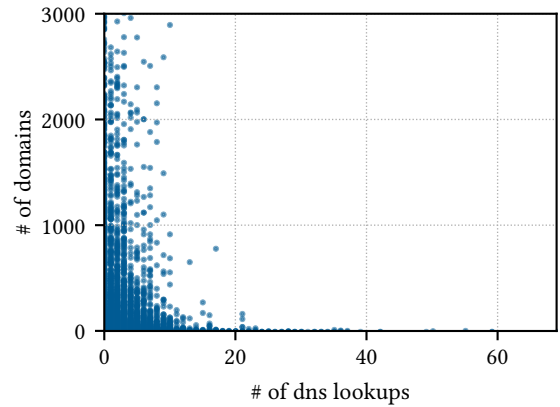}}
	\vspace{-6pt}
	\caption{Cutout of the number of domains using a specific include depending on the DNS lookup count.}
	\label{fig:include_lookups_zoom}
\end{figure}

Now that we have discussed this type of error, let us get back to the underlying causes.
As there is only a limited set of mechanisms that could create these errors, we find that the \spfcode{include} mechanism is the main cause of this issue. 
Reasons for this include but are not limited to recommendations by email or web hosting providers.
As an example, \textit{bluehost.com}\footnote{\url{https://www.bluehost.com/}} is a provider that recommends customers to add an invalid SPF record that causes 14 DNS lookups.
In \autoref{fig:include_lookups_zoom} we see a scatter plot of the includes where each dot represents an include.
We zoom into the interesting part with more than 10 includes.
In total, there are \numprint{\IncludesGeTen} included SPF records exceeding the DNS lookup limit directly, affecting \numprint{\IncludesGeTenDomains} domains.
\numprint{\IncludesBluehostCom} (\IncludesBluehostComPerc) of them are from \textit{bluehost.com}.

\paragraph{Too many void DNS lookups}
This error is raised when there are two DNS errors during the evaluation of the SPF record.
DNS errors are empty results or NXDOMAIN.
With \SPFTooManyDNSLookupsPerc this error is less common.
Since this error refers to a DNS lookup limit like \emph{Too many DNS lookups} and the reasons for exceeding them are almost the same, we will refrain from explaining them in detail again.

\paragraph{Syntax Errors}
A more interesting group of errors are syntax errors.
These are caused by different oversights and shortcomings when creating an SPF record.
Common errors are typos, wrong mechanism names or concatenating up different DNS records.
With \SPFSyntaxErrorPerc it is the third-largest error group and the most diverse one in our study.

In our manual investigation, the first common mistake in this group is that mechanisms are misspelled.
We find that \SyntaxErrorIPvFourPerc (\numprint{\SyntaxErrorIPvFour}) of the syntax errors are using \spfcode{ipv4} instead of \spfcode{ip4} and \SyntaxErrorIPvSixPerc (\numprint{\SyntaxErrorIPvSix}) are using \spfcode{ipv6} instead of \spfcode{ip6}.
\SyntaxErrorIPPerc (\numprint{\SyntaxErrorIP}) are just using \spfcode{ip} as the wrong mechanism. 
Similarly, merging DNS entries also leads to errors.
We observe that \SyntaxErrorSiteVerificationPerc (\numprint{\SyntaxErrorSiteVerification}) of the errors are concatenations of the SPF record and a site verification string.
When measuring the appearances of \spfcode{v=spf1} in the SPF records, the result is that \SyntaxErrorMultipleSpfPerc (\numprint{\SyntaxErrorMultipleSpf}) of the records with invalid syntax contain more than one, which could be caused by combining multiple recommendations.
A mechanism can have an argument that is placed directly after a \spfcode{:}, however a whitespace in this position is causing \SyntaxErrorWhitespacePerc (\numprint{\SyntaxErrorWhitespace}) of the errors.
Even though this group of errors is more diverse, they are typically easier to fix than other errors in SPF entries.

\paragraph{Include loops}
An include loop is created when an \spfcode{include} mechanism refers back to itself, either directly or at a deeper level of recursion.
It is a less common mistake with \SPFIncludeLoopPerc.
A direct inclusion of the domain happens in \SPFIncludeLoopOnePerc (\numprint{\SPFIncludeLoopOne}) of the cases.
We assume that knowledge about SPF is not correct in these cases.
When the error occurs at lower recursion levels, it is not obvious to detect and the cause is more intelligible.

\paragraph{Redirect loops}
Loops can also occur with \spfcode{redirect} mechanisms, representing \SPFRedirectLoopPerc of the errors.
The causes are similar to the \emph{include loops}.

\paragraph{Invalid IP address}
Because IP addresses have a well-defined representation, they can be easily written incorrectly. In our analysis,  this issue causes \SPFInvalidIPPerc of all errors. In particular, we observe the following four types of errors:%
\begin{itemize}
	\item No IP at all
	\item Wrong number of octets
	\item A domain instead of IP address
	\item Wrong IP version
\end{itemize}

\smallskip
\noindent %
Overall, our analysis of the errors shows that one of their main causes is insufficient attention to detail when creating SPF entries. Although SPF is a simple mechanism, the development of configurations is non-trivial and sometimes fraught with small details. For example, DNS lookup limits are challenging to inspect, inclusions and redirections can cause different loops, and the syntax of some SPF mechanisms must also be carefully considered.

\subsection{Notification}

Our detailed analysis of errors puts us in a unique position: We become able to run a notification campaign, informing domain operators about the discovered problems in their SPF configurations. To this end, we follow the recommendations developed by \citet{Stock2016, Stock2018} for large-scale notification and contact each operator via email. To reach as many operators as possible, we use the general addresses \spfcode{postmaster@} and \spfcode{security@} as defined in RFC2142 \cite{rfc2142} for our campaign. In addition, we send an email to the contact named in \spfcode{security.txt}~\citep{rfc9116}, if available.

\paragraph{Sending out notifications.}
In total, we sent \numprint{\NotificationsCount} mails to notify domain operators with erroneous SPF records, except for record-not-found errors.
We used a dedicated email server to deliver these huge amounts of emails.
To avoid being blacklisted, we throttled the transfer rate to 1 mail per second. Based on this limit, we sent out all notifications in the second week of May 2023.

For each notification email, we follow a fixed template: First, we introduce ourselves and the scope of our study. Then, we list the identified problems for the particular domain, along with examples and recommendations on how to fix them. We are aware that our campaign causes additional work for the operators, and therefore strive to provide actionable items for each error type. Further details on ethical considerations arising from this notification campaign are discussed in Appendix~\ref{app:ethics}.

\paragraph{Returned emails and feedback.}
It is clear that a notification campaign targeting hundreds of thousands of domains results in a large number of bounces and error messages. Nevertheless, we obtained a notable amount of positive feedback with thank-you notes, further questions and recommendations for future activities. By the time of the paper submission, we had received \numprint{\NotificationAnswersThx} grateful emails from domain operators. Only 3 responses were negative, and considered our notifications to be spam. We added the respective domains to an opt-out list so that they would not receive further security notifications from us.

\paragraph{Impact of notification}

To learn about the practical impact of our notification campaign, we rescanned the domains with errors on May 24, 2023, that is, two weeks after the notification.
We observe that \numprint{\SPFRescanResolvedErrors} errors have been fixed by that time. In the same period \numprint{\RescanEmptyDomains} of the domains with errors disappeared, so there are no errors anymore.
\autoref{tab:notification_success} shows detailed results for the different errors. The highest success rate is achieved with syntax errors and invalid IP addresses, as these can be easily fixed and do not require a deep understanding of SPF record evaluation. The errors with the lowest success rate are those related to DNS lookup limits. We assume that these are often non-trivial to fix, as they depend on the inclusion of external providers in the respective SPF configurations.

\begin{table}[htbp]
\begin{threeparttable}
	\centering
	\caption{SPF errors before and after our notification. }
	\label{tab:notification_success} \small
	\begin{tabular}{lrrr}
\toprule
\textbf{Error} & \textbf{Before} & \textbf{After} & \textbf{Change} \\
\midrule
Syntax Error & \numprint{38296} & \numprint{36103} & -5.73~\% \\
Too Many DNS Lookups & \numprint{49421} & \numprint{48630} & -1.60~\% \\
Too Many Void DNS Lookups & \numprint{5308} & \numprint{5127} & -3.41~\% \\
Redirect Loop & \numprint{58} & \numprint{56} & -3.45~\% \\
Include Loop & \numprint{19356} & \numprint{18617} & -3.82~\% \\
Invalid IP address & \numprint{7882} & \numprint{7498} & -4.87~\% \\
\midrule
Total Errors & \numprint{211018} & \numprint{204087} & -3.28~\% \\
\bottomrule
\end{tabular}

 \end{threeparttable}
\end{table}

Overall, our campaign achieves similar performance to notifications performed in previous work.
\citet{Stock2016}, for example,  report a 4.1\% success rate in reporting web vulnerabilities via email. Our campaign achieves a success rate of \NotificationSuccess just two weeks after sending the notifications, thus providing a similar effectiveness.

\subsection{Additional Findings}

We conclude our examination of SPF configurations with a discussion of further and curious findings discovered during the processing of our dataset.

\paragraph{Permissive all policies}
For \NoDenyAllPerc (\numprint{\NoDenyAll}) of the domains, the SPF configuration is missing a restrictive all policy, which harms the effectiveness of SPF.
The SPF evaluation will then just end without a matching mechanism and therefore return a \spfcode{neutral} result.
This may be intentional, as we will present in \autoref{sec:huge_cidr} for a few domains, but it leads to a reduction in protection.
In most cases, we notice that a final deny directive is missing, such as \spfcode{-all}.
Here, we often spot typos as the reason for the problem, such as the invalid terms \spfcode{-al} or \spfcode{-all;} in the SPF entries.

\paragraph{Not recommended records}
Over time, the SPF extension has evolved from the experimental RFC4408~\cite{rfc4408} to a proposed standard in RFC7208~\cite{rfc7208}.
Due to this evolution, the DNS record type \spfcode{SPF} has been deprecated since 2014.
The \spfcode{PTR} mechanism is not recommended anymore, as it is slow, not reliable due to DNS errors and produces a high DNS load. 
In our dataset, we find \numprint{\DeprecatedSPFDomains} domains still using this DNS record type and \numprint{\DeprecatedPTRDomains} domains using the \spfcode{PTR} mechanism. As these versions still provide protection, we do not count them as errors in our analysis.

\paragraph{Implementation of abuse reporting}
With RFC6652~\citep{rfc6652}, SPF was extended in 2012 with three new modifiers: \spfcode{ra}, \spfcode{rp} and \spfcode{rr}. These modifiers allow the operator of a domain to be notified when an unauthorized email is rejected at an email server. Although this is a helpful extension, we notice only \numprint{\ReportMechanism} domains implementing it in our dataset.

\paragraph{XSS attacks over SPF}
Finally, we observe a cross-site scripting attack packaged in an SPF record of a domain. The attack looks as follows:
\begin{center}
\color{cbone}
\begin{verbatim}
 v=spf1 xss=<script>alert('SPF')</script> ~all
\end{verbatim}
\end{center}
Since SPF parsers in email servers generally do not interpret JavaScript code, they should not be vulnerable to this type of attack.
However, as soon as software displays SPF records in a web browser, there is a risk that the attack will succeed.
This is, for example, the case for web services that check and validate SPF entries. Given the harmless payload of the attack, however, we assume that it is meant for testing purposes.

\section{SPF includes and spoofing}
\label{sec:includes}

During our analysis of \ac{SPF} entries, we observe several entries that authorize a very large number of IP addresses. In the following, we analyze these lax configurations in detail, investigate the underlying reasons and outline attacks that become possible through such configurations.

\subsection{Number of Authorized IP Addresses}
Since the goal of \ac{SPF} is to authorize senders for a given domain, the number of allowed sending IP addresses should be minimal to reduce the attack surface.
While the actual number of sending hosts for a domain is generally unknown, we can use the number of receiving servers to get at least an intuition of the general magnitude.
\citet{Ruohonen2020} reports that domains listed in the Alexa rankings generally have fewer than 20 MX records.
Therefore, we conjecture that the scale of sending servers is not significantly larger.
However, we find that many domains in our study authorize \emph{orders of magnitude} more addresses to send emails.

\begin{figure}[htbp]
	\centerline{\input{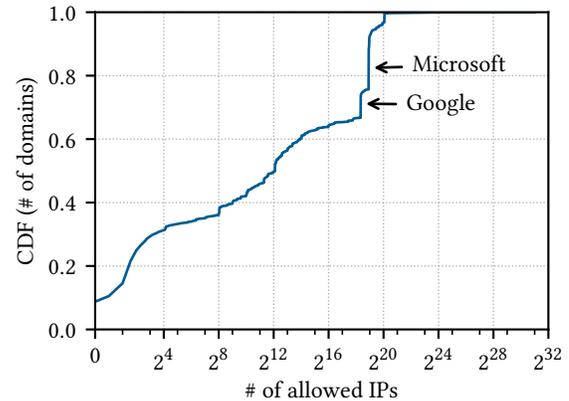}}
	\vspace{-6pt}
	\caption{CDF of authorized IPv4 addresses.}
	\label{fig:allowedip4}
\end{figure}

\autoref{fig:allowedip4} shows the distribution of the number of allowed IPv4 addresses as \ac{CDF}.
In line with our assumption, one out of three domains has fewer than 20 allowed hosts for sending emails.
By contrast, we also find that almost the same number of domains authorizes more than \numprint{100000} IPv4 addresses.
The largest rise in the \ac{CDF} is between \numprint{400000} and \numprint{700000} IPv4 addresses, mainly caused by including huge providers.
In general, there are two ways a huge number can arise in the SPF record.
First, we want to look at large IP ranges and later at includes. 

\subsection{Large IP Ranges}
\label{sec:huge_cidr}
In general, the reason for intentionally allowing large IP ranges is not clear to us.
To investigate, we take a look at SPF records that have more than \numprint{100000} IP addresses allowed.
Possible mechanisms are \spfcode{a}, \spfcode{mx} and \spfcode{ip4}.
We observed that \numprint{\HugeNoIncludeDomains} domains have their large number of IP addresses through these mechanisms.

\begin{table}[htbp]
	\centering
	\caption{Type and amount of SPF mechanisms with large IP ranges.}
	\label{tab:huge_cidr} \small
	\begin{tabular}{lrr}
\toprule
\textbf{CIDR} & \textbf{SPF:} 
\spfcode{ip4}, \spfcode{a}, \spfcode{mx} 
& \textbf{SPF:} \spfcode{include} \\
\midrule
/0 & \numprint{54} & \numprint{0} \\
/1 & \numprint{29} & \numprint{2} \\
/2 & \numprint{47} & \numprint{10} \\
/3 & \numprint{16} & \numprint{7} \\
/4 & \numprint{7} & \numprint{3} \\
/5 & \numprint{6} & \numprint{0} \\
/6 & \numprint{4} & \numprint{0} \\
/7 & \numprint{4} & \numprint{0} \\
/8 & \numprint{2162} & \numprint{110} \\
/9 & \numprint{23} & \numprint{3} \\
/10 & \numprint{131} & \numprint{27} \\
/11 & \numprint{44} & \numprint{50} \\
/12 & \numprint{313} & \numprint{137} \\
/13 & \numprint{228} & \numprint{210} \\
/14 & \numprint{1178} & \numprint{5419} \\
/15 & \numprint{1145} & \numprint{5389} \\
/16 & \numprint{11126} & \numprint{14243} \\
\bottomrule
\end{tabular}

\end{table}

In \autoref{tab:huge_cidr} we see an overview about the appearance of very large IP ranges.
At the hugest possible network \spfcode{/0} we found \numprint{\AllowInternet} domains explicit allowing \spfcode{0.0.0.0/0}, what looks intentional.
In contrast, there are \FPeval{\result}{round(\DomainsCidrZero-\AllowInternet,0)}\result{} domains that have a specific IPv4 address with a tailing \spfcode{/0}, what rather appears to be a misunderstanding of CIDR prefixes.
Going ahead, the huge includes \spfcode{/1} and \spfcode{/2} appear to be typos that should refer to \spfcode{/16} and \spfcode{/24} respectively.
Continuing the rows in \autoref{tab:huge_cidr} we can see that the includes are lower than the direct mechanisms, both at a very low level. 

Therefore, these few large IP ranges in SPF Records cannot explain the huge amount of allowed IP addresses we see in Figure 5 around $2^{19}$.
In the next section, we will have a detailed look at the \spfcode{include} mechanism and its impact on the number of allowed IP addresses.
This is more promising to be the reason, as we observe that \numprint{\HugeIncludeDomains} domains authorize a large number of IP addresses through the \spfcode{include} mechanism.

\subsection{Usage of Includes}
To get a better understanding of why so many IP addresses are included in the \ac{SPF} entries, we analyze the use of the \spfcode{include} mechanism. 
This mechanism is designed to cross administrative borders and is used by \IncludeDomainsPerc of the domains.
Providers often recommend to their customers that they add a specific \spfcode{include} mechanism to their \ac{SPF} record when using their services.
As already mentioned, \numprint{\HugeIncludeDomains} domains have a coarse policy with a huge number of allowed IP addresses from an include.

\begin{figure}[htbp]
	\centerline{\input{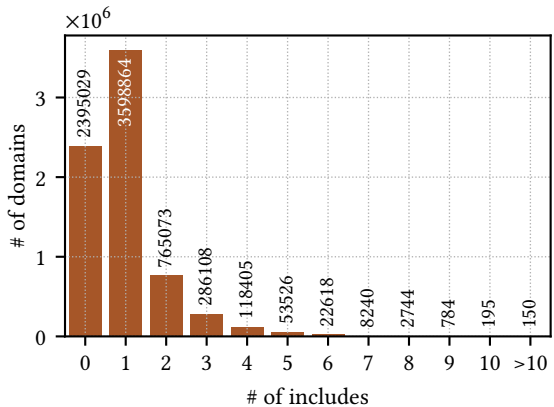}}
	\vspace{-6pt}
	\caption{Number of includes in the top level record.}
	\label{fig:toplevel_includes_valid}
\end{figure}

\paragraph{Trust relationships.}
In general, including a configuration from another domain involves a certain trust related risk.
The owner has no control over possible changes in the inherited addresses, which could lead to spoofed email addresses with valid \ac{SPF} check.
Therefore, the domain owner has to trust the party they include from.
While it should not be a big problem to trust the own provider in this case, things change if there are multiple inclusion levels and thus several administrators involved.
As shown in \autoref{fig:toplevel_includes_valid}, most configurations have not more than one include, which seems reasonable in the described scenario.
Nevertheless, we also observed 10 recursive includes and more, which raises the question if the domain owners are aware of everything they include and trust all involved parties.

\begin{figure}[htbp]
	\centerline{\input{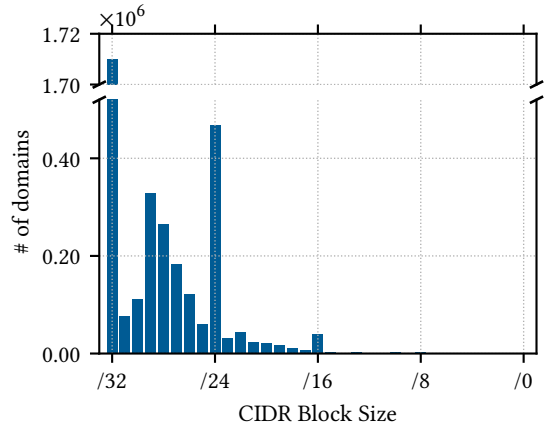}}
	\vspace{-6pt}
	\caption{Distribution of subnet sizes in includes.}
	\label{fig:subnet_size_valid}
\end{figure}

\paragraph{Included network size.}
Another interesting part of the includes are the networks allowed by them.
\autoref{fig:subnet_size_valid} shows the distribution of the used network sizes coming from the included \ac{SPF} records.
While most entries only include one IP address (/32 network), there is also a second notable peak for /24 networks.
In the context of larger providers, load balancing and scaling, these sizes are understandable to share the load between multiple servers.
Surprisingly, there are also \ac{SPF} entries, which allow very huge networks, larger than /16.
Even though large providers like Google might need a large number of servers sending mail for their customers, there are obviously limits.
We could not find a specific reason for these includes.
Especially for less common domains, at the end of the Tranco list, we observe many /8 inclusions.
Malicious senders could use these domains to send emails from many hosts within this list.
What is interesting here is that most of the domains we observe in this context come from the ".top" top-level domain.  

\begin{figure}[b]
	\centerline{\input{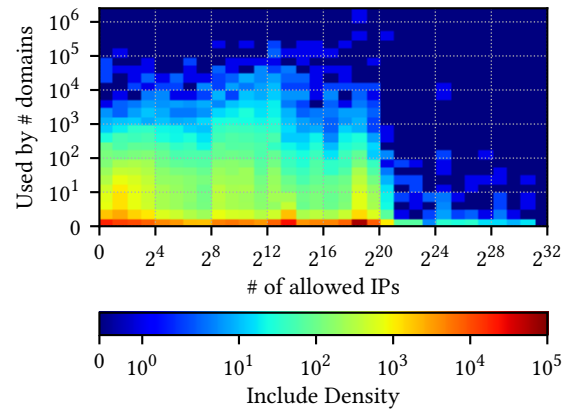}}
	\vspace{-6pt}
	\caption{Heatmap of domains that are using a specific include, depending on the allowed IPs for the include.}
	\label{fig:include_usage_all}
\end{figure}

\paragraph{Number of IP addresses per include.}
We now take a closer look at how many times an \spfcode{include} is used, depending on the number of allowed IP addresses.
\autoref{fig:include_usage_all} is a heatmap that shows the density of includes within a pixel of the plot, representing a logarithmic scale of the number of allowed IPs and how often the include is used.
We can see that there is a huge concentration, up to around $2^{20}$ allowed IPs.
Recalling \autoref{fig:allowedip4}, we see that this correlates with the steep rise there at the same number of domains. From this observation, we can conclude that the large numbers of allowed IPs are typically from includes.

In \autoref{tab:top20_includes}, we report the top 20 includes we discover in our scan.
As the first two, namely Microsoft and Google, include a huge amount of IP addresses. Similarly, other includes of providers are larger than \numprint{100000} and likely too coarse for proper protection. In contrast, there are also providers, like OVH\footnote{\url{https://www.ovhcloud.com}} and Xserver\footnote{\url{https://www.xserver.ne.jp/}}, which include only a few sending email servers, demonstrating a restrictive authorization policy. 

\begin{table}[htbp]
\begin{threeparttable}
	\centering
	\caption{Top 20 included domains with their number of allowed IPs.}
	\label{tab:top20_includes} \small
	\begin{tabular}{lrr}
\toprule
\textbf{Include} & \textbf{Used by} & \textbf{Allowed IPs} \\
\midrule
\detokenize{spf.protection.outlook.com} & \numprint{2456916} & \numprint{491520} \\
\detokenize{_spf.google.com} & \numprint{1418705} & \numprint{328960} \\
\detokenize{websitewelcome.com} & \numprint{414695} & \numprint{1088784} \\
\detokenize{secureserver.net} & \numprint{374986} & \numprint{505104} \\
\detokenize{relay.mailchannels.net} & \numprint{289112} & \numprint{4358} \\
\detokenize{servers.mcsv.net} & \numprint{263343} & \numprint{22528} \\
\detokenize{spf.mandrillapp.com} & \numprint{236293} & \numprint{4608} \\
\detokenize{sendgrid.net} & \numprint{215497} & \numprint{220672} \\
\detokenize{_spf.mailspamprotection.com} & \numprint{212418} & \numprint{1049} \\
\detokenize{spf.efwd.registrar-servers.com} & \numprint{196465} & \numprint{264} \\
\detokenize{amazonses.com} & \numprint{183184} & \numprint{64512} \\
\detokenize{mx.ovh.com}\tnote{1} & \numprint{176191} & \numprint{2} \\
\detokenize{mailgun.org} & \numprint{172499} & \numprint{36312} \\
\detokenize{_spf.mail.hostinger.com} & \numprint{139423} & \numprint{4358} \\
\detokenize{zoho.com} & \numprint{138227} & \numprint{6209} \\
\detokenize{mail.zendesk.com} & \numprint{114026} & \numprint{26112} \\
\detokenize{spf.mailjet.com} & \numprint{111760} & \numprint{5120} \\
\detokenize{spf.web-hosting.com} & \numprint{111405} & \numprint{10492} \\
\detokenize{spf.sendinblue.com} & \numprint{102004} & \numprint{87040} \\
\detokenize{spf.sender.xserver.jp} & \numprint{92411} & \numprint{15} \\
\bottomrule
\end{tabular}

    \begin{tablenotes}
        \item[1] Uses not recommended PTR mechanism
    \end{tablenotes}
\end{threeparttable}
\end{table}

\subsection{Case Study on Web Hosting}

Our analysis shows that coarse authorization is a common practice in SPF configurations. 
From a theoretical point of view, it is obvious that overly permissive policies weaken the intended protection. 
However, whether these loose configurations can really help attackers in practice is not immediately clear.
We therefore set out to investigate the risk of this practice in a case study on web hosting providers.

In particular, we focus on common web hosting providers that usually offer web space along with email support.
Since these providers manage thousands of domains for their customers, they represent an essential part of the Tranco list and the collected SPF records.
Moreover, as most web hosting providers support active content, such as PHP scripts, we are able to test the sending and authorization of SPF with permissive configurations.

As the basis for the case study, we search for recommended providers worldwide using the review website \emph{hostings.info}\footnote{\url{https://hostings.info/}}.
From each country in the overview, we search for a recommended SPF record at the top 10 recommended web hosters.
Due to the size of our measurements, we found 79 providers.
Since many of the providers require national residency or the purchase of a national domain, we rent web space from 5 providers that do not impose any constraints.
We especially choose providers that offer PHP support, allow the use of own domains, and provide a short contract period.
These providers are located in 4 countries (2xDE, FR, US, UK).

\paragraph{Imitating spoofing}
To investigate the risk of spoofing due to lax SPF configuration, we sent emails from the selected providers to ourselves.
In these emails, we spoof the sending domain by picking one that authorizes the IP address of the web hosting provider due to the recommended SPF record.
Technically, we use two methods to realize this strategy: First, we try to send an email directly via SMTP from the web space via a corresponding PHP script.
Second, we use the PHP function \spfcode{mail()} to send it via the local \ac{MTA} of the provider.
We then examine how the emails are received on our site and whether they pass the SPF checks.
A spoofing attempt is considered successful if one of the two methods succeeds in transferring a valid email.

Note that we only send emails from our rented web space to our own email addresses.
Thus, spoofed senders in these mails do not cause any harm.
Also, we have informed the vulnerable web hosting providers about their lax configurations in the hope that they will enforce stricter policies.
For more details on ethical considerations, see \autoref{app:ethics}.

\paragraph{Results}
We find that 4 of the 5 web hosting providers enable us to send emails with spoofed senders due to overly coarse authorization, as we can see in \autoref{tab:provider_results}.
In particular, there are two providers that enable sending emails with the PHP \spfcode{mail()} function over their \ac{MTA}.
Among these two providers, we find \numprint{264} and \numprint{24959} authorized domains, respectively.
In addition, there is one provider that allows us to send emails via SMTP for \numprint{159} affected domains.
With the fourth provider, we can even send emails via the SMTP method and \spfcode{mail()} on behalf of \numprint{713} domains.

In summary, we are able to send emails with valid SPF entries from \numprint{26095} domains, simply by renting web hosting space for about 30 Euro. Even worse, we can pick from a wide range of domains for the spoofing, including lobby organizations, political parties, health insurances companies, and even banks. Although our case study focuses on a small group of web hosting providers, it shows the potential for phishing campaigns and spam when lax SPF configurations are exploited by attackers.
\begin{table}[htbp]
	\centering
	\caption{Results of the providers case study.}
	\label{tab:provider_results}
\begin{tabular}{@{}ccrr@{}}
\toprule
\textbf{Provider} & \textbf{Success} & \textbf{\# Domains} & \textbf{\# Allowed IPs} \\ \midrule
1                 & MTA              & \numprint{24959}   & \numprint{177168}      \\
2                 & SMTP, MTA        & \numprint{713}     & \numprint{514}         \\
3                 & MTA              & \numprint{264}     & \numprint{2052}        \\
4                 & SMTP             & \numprint{159}     & \numprint{3074}        \\
5                 & None             & \numprint{0}       & \numprint{672}         \\ \bottomrule
\end{tabular}
\end{table}

\section{Lesson Learned}
\label{sec:recommendations}

After examining the prevalence of flaws in SPF implementations, we work out some recommendations on how to improve the use of SPF. We first discuss possible actions for domain owners before we take a look at the side of the web hosting providers.

\subsection{Domain Owners}
\label{sec:recommendations_domain}
Domain owners are often dependent on other parties to operate particular services.
This especially applies to email servers, where an external provider is often responsible for the security of the service.
Nevertheless, in most cases, the domain owner needs to take care of all DNS records and therefore must provide the correct SPF configuration.

If the email server is operated by another provider, the domain owner should, in general, follow their recommendations for \ac{SPF} records.
As they might change IP addresses from time to time, this is usually a scenario for which the \spfcode{include} mechanism is intended.
However, we recommend checking the included addresses.
As can be seen from our analysis, often only a single inclusion is necessary, making such a test technically feasible.

On the other hand, one of our findings is that providers may recommend including large IP ranges or additional includes in order to use their service.
In these cases, we strongly recommend to check if the included ranges are only email servers used for the domain, potentially by contacting the provider and requesting a description of the includes.
A further risk is an \spfcode{a} mechanism in the SPF record of a shared web space.
Every user on the same server that the A record points to could use this server to send an email on behalf of this domain.
Ultimately, the recommended SPF entry can be taken as a rough indicator of whether a provider takes email security seriously and therefore serves as a decision-making aid for choosing a provider.

If the domain owners manage their DNS records themselves, they are fully responsible for the content.
Our work shows that there are many entries with syntax errors, which could easily be prevented in advance.
We therefore recommend validating SPF records with a tool to check for errors and undefined parts.
In case of a self-hosted email infrastructure, administrators should ensure that they only add the hosts needed to send emails.

\subsection{Web Hosting Providers}
\label{sec:recommendations_provider}

Web hosting providers generally want to offer their customers a well-functioning and user-friendly service, yet this sometimes conflicts with providing the best possible security.
A well-designed setup can help avoid difficulties.

Customers should usually not be able to open \ac{SMTP} connections directly to email servers on a shared web space, especially if this host is included in the recommended \ac{SPF} entry.
Therefore, it is a recommended practice to block outgoing connections to port 25 and related services.
Nevertheless, users should be able to send emails within their application.
Therefore, a local \acp{MTA} with proper authentication should be used to verify that the authenticated account is allowed to send emails on behalf of the specified domain.

If a customer needs to send emails directly using their own \ac{MTA}, this should not be done using shared IP addresses.
We recommend providing a user documentation to manually add an IP address to the entry, instead of automatically including it in the default \ac{SPF} record.
To prevent further problems with \ac{SPF}, providers should enable their customers to understand and properly use this framework.
Therefore, they should explain how it works or link to relevant material, and also point out potential risks associated with setting certain SPF mechanisms. 

\section{Conclusions}
\label{sec:conclusions}

With our analysis, we shed light on the state of SPF in the wild.
We observe an increasing adoption of this security mechanism; at the same time, we find flawed and overly coarse authorization policies in numerous cases.
We demonstrate that these lax practices increase the attack surface of SPF and make spoofing senders possible with little effort.
It is enough to identify web hosting providers that manage thousands of domains with permissive configurations to send spoofed emails at a large scale.

Fortunately, we can conclude from our notification campaign that several of the configurations were not intentionally malfunctioning.
Shortly after our notifications, we could already observe thousands of fixed SPF entries. In general, SPF faces a tradeoff between security and usability.
Although a minimal authorization policy would be desirable, operators often relax their configurations for practical reasons, for instance, because it is inconvenient to identify all sending hosts or because they try not to interfere with their clients' activities.
Our analysis shows that the compromises made by operators are far from adequate, and we therefore strongly recommend using more validated and restrictive SPF policies in practice, for example, by following the guidelines presented in this paper.

\section*{Acknowledgements}
We would like to thank our shepherd Anna Sperotto and the anonymous reviewers for their valuable comments and suggestions.
We also thank Frank Rust from TU Braunschweig for his extraordinary support and Mike Cardwell for coming up with XSS attacks in SPF entries.
This work was funded by the German Federal Ministry of Education and Research under the grant BIFOLD23B, and the European Research Council (ERC) under the consolidator grant MALFOY (101043410).

\balance
\bibliographystyle{abbrvnat}
\bibliography{bibliography}

\appendix

\section{Ethics}
\label{app:ethics}
Our university does not implement a formal IRB process for the conducted study. Still, we have designed all experiments in accordance with ethical best practices outlined in the Menlo report~\citep{KenDit12} and legal regulations of the European GDPR~\citep{GDPR}. First, the collection of SPF and DMARC records is fully automated and does not involve human subjects. By design, the collected data is openly available and does not contain any private or sensitive information. Furthermore, we have taken measures to keep the load on DNS servers as low as possible.
To this end, we implemented a cache as described in \autoref{sec:measuring-in-the-wild}.
Second, we have notified all operators of domains with misconfigured SPF records via email, providing detailed descriptions of the identified issues.
Although this notification caused extra work for the operators, we argue that informing them about the misconfigurations and thereby improving email protection outweighs this disadvantage.

\end{document}